\documentclass{article}
\usepackage[T1]{fontenc}

\usepackage{graphicx}
  
\usepackage[cmex10]{amsmath}
\usepackage{amssymb}
\usepackage{natbib}
\usepackage{authblk}

\usepackage{algorithmic}

\allowdisplaybreaks

\newtheorem{proposition}{Proposition}

\begin{document}

\title{Covariate Selection in High-Dimensional Generalized Linear Models With Measurement Error}

\author[1]{\O ystein S\o rensen\thanks{oystein.sorensen@medisin.uio.no}}
\author[1,2]{Arnoldo Frigessi\thanks{arnoldo.frigessi@medisin.uio.no}}
\author[1]{Magne Thoresen\thanks{magne.thoresen@medisin.uio.no}}

\affil[1]{Oslo Centre for Biostatistics and Epidemiology, Department of Biostatistics, University of Oslo, Norway}
\affil[2]{Oslo Centre for Biostatistics and Epidemiology, Research Support Services, Oslo University Hospital, Norway}

\maketitle

\begin{abstract}
In many problems involving generalized linear models, the covariates are subject to measurement error. When the number of covariates $p$ exceeds the sample size $n$, regularized methods like the lasso or Dantzig selector are required. Several recent papers have studied methods which correct for measurement error in the lasso or Dantzig selector for linear models in the $p > n$ setting. We study a correction for generalized linear models based on Rosenbaum and Tsybakov's matrix uncertainty selector. By not requiring an estimate of the measurement error covariance matrix, this generalized matrix uncertainty selector has a great practical advantage in problems involving high-dimensional data. We further derive an alternative method based on the lasso, and develop efficient algorithms for both methods. In our simulation studies of logistic and Poisson regression with measurement error, the proposed methods outperform the standard lasso and Dantzig selector with respect to covariate selection, by reducing the number of false positives considerably. We also consider classification of patients on the basis of gene expression data with noisy measurements.
\end{abstract}


\maketitle

\section{Introduction}

Regularization methods like the lasso \citep{Tibshirani96} and Dantzig selector (DS) \citep{Candes07} are widely used in regression, when the number of covariates $p$ exceeds the number of measurements $n$. Lasso and DS typically select a small set of covariates which are given nonzero estimated regression coefficients. In addition to yielding models with good predictive performance, the set of selected covariates can be a good starting point for further scientific investigations. For example, in classification of cancer patients based on gene expression measurements, the selected genes can be interesting candidates for further study. The covariate selection and parameter estimation properties of the lasso and the closely related DS have received much attention in the literature, e.g., \cite{Bickel09,Meinshausen2006,vandegeer2009,Wainwright2009,Zhao06}, summarized in the recent monograph \cite{Buhlmann11}. Bounding the estimation error of the regression coefficients with high probability requires that the covariates satisfy a restricted eigenvalue condition \citep{Bickel09}, while perfectly recovering the set of nonzero coefficients with the lasso requires a stronger irrepresentable condition \citep{Meinshausen2006,Zou2006,Zhao06}. 

Both the lasso and the DS can be defined for generalized linear models (GLMs) \citep{McCullagh89}. The framework of GLMs includes logistic regression for binomial outcomes and Poisson regression for outcomes which can be modeled by a Poisson process. The extension of lasso to GLMs was derived in the original lasso paper \citep{Tibshirani96}, while the generalized DS (GDS) has been proposed by \cite{James09}. For the lasso, the theoretical results for linear regression models have been extended to GLMs \citep{Buhlmann11,vdGeer2008}.

An implicit assumption in all the work mentioned so far, is that the covariates are perfectly measured, i.e., that the only source of noise is the stochastic relationship between the measured covariates and the outcome. In reality, most problems are subject to at least a small amount of measurement error. For example, gene expression microarray measurements are subject to various sources of systematic and random error \citep{Rocke01}, and are a noisy version of the true gene expression in the patients. Another example is food frequency questionnaires (FFQs) used in epidemiologic studies, in which subjects are asked about their food consumption. It is well known that the responses given in FFQs are often far away from the true food consumption \citep{Kipnis2003}. Sensor network data also tend to be noisy, due to measurement error or sensor failure \citep{Bertrand2011}. 

In classical regression problem ($p<n$), measurement error is known to yield biased parameter estimates and lack of power \citep{Carroll06}. Correction methods typically require replicate measurements of the covariates in order to estimate the measurement error distribution. We refer to the excellent monograph \cite{Carroll06} for an overview of measurement error models in statistics. Correction for measurement error in penalized regression has been studied by various authors recently \citep{Chen2013,Liang2009,Loh12,Ma2010,Rosenbaum10,Rosenbaum13,Sorensen13,Zhu11}. Other authors have considered measurement error in the response rather than in the covariates, e.g., \cite{Nguyen2013}. Under a sparsity assumption, it has been shown that the standard formulations of the lasso and DS may select too many covariates in the presence of measurement error \citep{Rosenbaum10,Sorensen13}. The correction methods studied by \cite{Liang2009,Loh12,Ma2010,Rosenbaum13,Sorensen13} require knowledge of the measurement error distribution. These likelihood based approaches yield estimators with good statistical properties. However, in many applications with large $p$, it may be hard even to obtain a rough estimate of the $p\times p$ covariance matrix of the measurement error, which would be needed in the case of Gaussian additive errors. In contrast, the matrix uncertainty selector (MUS) of \cite{Rosenbaum10}, the sparse total least squares (S-TLS) method of \cite{Zhu11}, and the orthogonal matching pursuit (OMP) algorithm of \cite{Chen2013}, account for the measurement error without requiring an estimate of its distribution. The latter methods will hence have a practical advantage in many applications, and in particular, they have been shown to yield fewer false positive (FP) selections than the standard lasso and DS.

The methods proposed for dealing with measurement error in penalized regression all focus on linear regression, with the exception of \cite{Ma2010} and \cite{Sorensen13}. Considering the importance and general applicability of GLMs, it is therefore of interest to develop penalized regression methods for GLMs which do not require an esimate of the measurement error distribution, and recover the linear regression framework as a special case. In this paper, we propose the generalized MUS (GMUS), based on a Taylor expansion of the GLM mean function around the true, but unknown, covariates. The GMUS can be computed using iteratively reweighted least squares (IRLS), and when the Taylor expansion is truncated at first order, each step of the IRLS algorithm requires solving a simple linear program. We also develop a generalized matrix uncertainty lasso (GMUL), which is a lasso-type analog of the DS-based GMUS. The GMUL estimate can also be computed using an IRLS algorithm, in which an inner coordinate descent loop has to be run until convergence in each step of the algorithm. In simulation experiments with logistic and Poisson regression, the GMUL and GMUS with the first order Taylor approximation are shown to give very promising covariate selection results compared to the lasso and the GDS, by detecting considerable fewer false positives (FPs) at similar numbers of true positives (TPs). 

The outline of the paper is as follows. In Section \ref{sec:Background} we define the setup and notation, and give a brief overview of the relevant background. In Section \ref{sec:MUS} we derive the GMUS for GLMs with measurement error as an extension of the MUS for linear models. We also describe an IRLS algorithm for computing the GMUS. Next, in Section \ref{sec:MUL} we show how a lasso analog of the GMUS can be defined through the KKT conditions, which we term the GMUL. An algorithm for computing the GMUL is developed, which requires an outer IRLS loop and an inner coordinate descent loop. In Section \ref{sec:Convergence} we investigate the convergence of our algorithms for the GMUL and the GMUS in a simple numerical experiment. In Section \ref{sec:Simulation} we present the results of simulation experiments comparing both methods to the standard formulations of the GDS and the lasso for logistic and Poisson regression. The simulation studies confirm earlier results showing that the standard methods grossly overestimate the number of nonzero regression coefficients in the presence of measurement error. In the settings considered, the GMUL and GMUS reduce the number of false positive selections considerably, without missing many true positives. Finally, in Section \ref{sec:Example} we apply the GMUL to a problem of finding genes which are differentially expressed between patients with low vs. high bone mineral density. Mathematical derivations are given in the appendices.

\section{Background and Model Setup}\label{sec:Background}
We consider a GLM \citep{McCullagh89} with response $Y$ distributed according to
\begin{equation}
f_{Y}\left(y; \theta,\phi \right) = \exp\left\{\frac{y\theta - b\left( \theta\right)}{a\left( \phi \right)} + c\left(y,\phi\right) \right\},
\label{eq:GLMresponse}
\end{equation}
with linear predictor $\theta = \mathbf{x}^{T}\boldsymbol{\beta}^{0}$. The covariates are $\mathbf{x} \in \mathbb{R}^{p}$, the set of nonzero coefficients $S^{0}=\left\{j: \beta_{j}^{0}\neq 0\right\}$ has cardinality $s = \text{card}\left\{S^{0}\right\}$, and $\boldsymbol{\beta}^{0} \in \mathbb{R}^{p}$ is the true vector of regression coefficients. In order to accomodate an intercept term, we implicitly assume that the first element of $\mathbf{x}$ is a constant $1$. The expected response is given by the mean function
\begin{align*}
\mu\left(\theta\right) = b'(\theta)= g^{-1}\left( \theta\right),
\end{align*}
where $g^{-1}\left(\cdot\right)$ is the inverse of a canonical link function $g\left(\cdot\right)$. For logistic regression, the mean function equals $\left(1+\exp\left(-\theta\right)\right)^{-1}$, and for Poisson regression it equals $\exp\left(\theta\right)$. We focus on cases in which the dispersion parameter $\phi$ is known or a constant, as it is for Poisson and logistic regression, and will hence neglect $a(\phi)$ for notational convenience in the sequel.

Given a data set with covariates $\mathbf{X} = \left(\mathbf{x}_{1},\dots,\mathbf{x}_{n}\right)^{T}$, and response $\mathbf{y} = \left(y_{1},\dots,y_{n}\right)^{T}$ distributed according to (\ref{eq:GLMresponse}), the lasso \citep{Tibshirani96} estimates $\boldsymbol{\beta}^{0}$ by maximizing the log-likelihood subject to an $\ell_{1}$-constraint on the regression coefficients. This yields a nonsmooth convex optimization problem, whose Lagrangian version is
\begin{equation}
\hat{\boldsymbol{\beta}}_{L}
\label{eq:LassoGLM} \in \text{arg}~\underset{\boldsymbol{\beta}}{\text{min}}\frac{-1}{n}\sum_{i=1}^{n}\left\{y_{i}\mathbf{x}_{i}^{T} \boldsymbol{\beta} - b\left(\mathbf{x}_{i}^{T}\boldsymbol{\beta} \right) \right\}+ \lambda \left\|\boldsymbol{\beta}\right\|_{1}
\end{equation}
where $\lambda$ is a regularization parameter. An alternative estimator is the GDS, 
\begin{align}\label{eq:GDS}
&\hat{\boldsymbol{\beta}}_{DS} \in  \text{arg}~\underset{\boldsymbol{\beta}}{\text{min}} \left\|\boldsymbol{\beta}\right\|_{1}, \text{ subject to (s.t.) }  \\ \nonumber
&\frac{1}{n}\underset{1\leq j \leq p}{\text{max}}\left| \sum_{i=1}^{n}x_{ij} \left\{y_{i} - \mu\left(\mathbf{x}_{i}^{T}\boldsymbol{\beta} \right) \right\}\right|\leq \lambda
\end{align}
which can be computed by iterative linear programming using an iteratively reweighted least squares (IRLS) approach \citep{James09}. The GDS and the lasso are closely related. For a given $\lambda$, we have
\begin{align*}
\hat{\boldsymbol{\beta}}_{DS},\hat{\boldsymbol{\beta}}_{L} \in 
\left\{ \boldsymbol{\beta} \in \mathbb{R}^{p}: \frac{1}{n}\underset{1\leq j \leq p}{\text{max}}\left| \sum_{i=1}^{n}x_{ij} \left\{y_{i} - \mu\left(\mathbf{x}_{i}^{T}\boldsymbol{\beta} \right) \right\}\right|\leq \lambda\right\},
\end{align*}
and hence, both for the lasso and for the GDS, the correlation of any covariate with the residual is bounded by $\lambda$. By definition, $\|\hat{\boldsymbol{\beta}}_{DS} \|_{1} \leq \|\hat{\boldsymbol{\beta}}_{L} \|_{1}$ \citep{Bickel09}.

\subsection{Covariate Measurement Error}
In (\ref{eq:LassoGLM}) and (\ref{eq:GDS}), it is implicitly assumed that the covariates are perfectly known. In many real applications, however, the measurement process is subject to noise and covariates may be missing at random. In particular, we consider the additive measurement error model
\begin{equation}\label{eq:AdditiveME}
\mathbf{w}_{i} = \mathbf{x}_{i} + \mathbf{u}_{i}, ~i=1,\dots,n,
\end{equation}
where $\mathbf{w}_{i}$ is the vector of measurements of sample $i$ and $\mathbf{u}_{i}$ is the vector of measurement errors. Problems with missing data can also be expressed in this additive form \citep{Rosenbaum10}. We further assume that the measurements in $\mathbf{w}_{i}$ are standardized to have zero mean and unit variance,
\begin{equation}\label{eq:StandardizedCovs}
\frac{1}{n}\sum_{i=1}^{n}w_{ij}=0 \text{ and } \frac{1}{n} \sum_{i=1}^{n} w_{ij}^{2} = 1, ~j=1,\dots,p,
\end{equation}
and let $\mathbf{W} = (w_{1},\dots,w_{n})^{T}$ and $\mathbf{U} = (u_{1},\dots,u_{n})^{T}$. Finally let
\begin{align*}
\epsilon_{i} = y_{i} - \mu\left(\mathbf{x}_{i}^{T}\boldsymbol{\beta}^{0}\right), ~ i=1,\dots,n
\end{align*}  
denote the generalized residual and define the vector $\boldsymbol{\epsilon} = (\epsilon_{1},\dots,\epsilon_{n})^{T}$.

In the absence of measurement error, $\boldsymbol{\beta}^{0}$ is contained in the feasible set of the GDS if $\lambda$ is chosen such that
\begin{equation}
\frac{1}{n}\left\|\mathbf{X}^{T}\boldsymbol{\epsilon} \right\|_{\infty} \leq \lambda
\label{eq:NoiseBoundNoMe}
\end{equation}
holds \citep{Antoniadis10,Candes07}. Hence, for any GDS solution $\hat{\boldsymbol{\beta}}_{DS}$ as well as for the true regression coefficients $\boldsymbol{\beta}^{0}$, the maximum correlation of any covariate with the residual is bounded by $\lambda$. Under restricted eigenvalue conditions, this also means that $\hat{\boldsymbol{\beta}}_{DS}$ and $\boldsymbol{\beta}^{0}$ are close, and that their maximum possible distance increases in $\lambda$. Hence, a natural starting point for a theoretical analysis is to assume that the bound (\ref{eq:NoiseBoundNoMe}) holds \citep[p. 103]{Buhlmann11}. However, when the true covariates are unknown, and noisy measurement $\mathbf{W}$ are plugged into the GDS, $\boldsymbol{\beta}^{0}$ is not guaranteed to be feasible even when (\ref{eq:NoiseBoundNoMe}) holds. The reason is that (\ref{eq:NoiseBoundNoMe}) only bounds the noise corresponding to the residual, whereas the noise in the measured covariates is not taken into account. For the special case of linear regression, \cite{Rosenbaum10} introduced a new parameter, $\delta$, which bounds the magnitude of the measurement error, yielding the two bounds
\begin{equation}\label{eq:MUBounds}
\frac{1}{n}\left\|\mathbf{W}^{T}\boldsymbol{\epsilon} \right\|_{\infty} \leq \lambda \text{ and } \left\|\mathbf{U}\right\|_{\infty} \leq \delta,
\end{equation}
where $\|\cdot\|_{\infty}$ is the maximum component norm.
When the bounds (\ref{eq:MUBounds}) hold, $\boldsymbol{\beta}^{0}$ is a feasible solution of the MUS, given by
\begin{align}\label{eq:MUS}
&\hat{\boldsymbol{\beta}}_{MU} \in \text{arg}~\underset{\boldsymbol{\beta}}{\text{min}} \left\|\boldsymbol{\beta}\right\|_{1}, \text{ s.t. }  \\ \nonumber
&\frac{1}{n}\left\|\mathbf{W}^{T}\left(\mathbf{y} - \mathbf{W}\boldsymbol{\beta}\right) \right\|_{\infty} \leq \lambda + \delta \left\|\boldsymbol{\beta}\right\|_{1}.
\end{align}
The MUS thus modifies the DS by adding the term $\delta \left\|\boldsymbol{\beta}\right\|_{1}$ to the upper bound on the correlation of the measurements with the residual, ensuring that $\boldsymbol{\beta}^{0}$ is still feasible. When $\delta$ is small, the MUS yields tight bounds for $\|\hat{\boldsymbol{\beta}}_{MU} - \boldsymbol{\beta}^{0}\|_{q}$, $q \in \{1,2\}$, as well as good covariate selection properties \citep{Rosenbaum10}. We finally note that the price to pay for not knowing the measurement error distribution, is that the estimation error bounds of the MUS do not go to zero when $n\to \infty$. However, if the main goal is covariate selection, the MUS may still have very good finite sample performance, as illustrated by \cite{Rosenbaum10}.

\section{MUS for High-Dimensional GLMs with Measurement Error}\label{sec:MUS}
We now present an extension of the MUS to GLMs with additive measurement error. In order to do this, we need to find a feasible set which contains $\boldsymbol{\beta}^{0}$ and yields good model fit, and then estimate $\boldsymbol{\beta}^{0}$ by a sparse element of this set.

First, consider a Taylor expansion of the mean function $\mu\left(\theta_{i} \right) = \mu\left(\mathbf{x}_{i}^{T}\boldsymbol{\beta}^{0}\right)$ around the scalar point $\mathbf{w}_{i}^{T}\boldsymbol{\beta}^{0}$,
\begin{equation}
\mu\left( \mathbf{x}_{i}^{T}\boldsymbol{\beta}^{0}\right) = \sum_{r=0}^{\infty} \frac{\mu^{\left(r\right)}\left(\mathbf{w}_{i}^{T}\boldsymbol{\beta}^{0} \right)}{r!}\left(-\mathbf{u}_{i}^{T}\boldsymbol{\beta}^{0}\right)^{r},
\label{eq:TaylorExpansion}
\end{equation}
where $\mu^{(r)}(\cdot)$ denotes the $r$th derivate of $\mu(\cdot)$. We now have the following result.
\begin{proposition}\label{prop:FeasibleSet}
Assuming the bounds (\ref{eq:MUBounds}) hold, the true regression coefficients satisfy $\boldsymbol{\beta}^{0} \in \Theta$, where 
\begin{align*}
 \Theta = \left[\boldsymbol{\beta}\in \mathbb{R}^{p}:\frac{1}{n}\underset{1\leq j \leq p}{\text{max}}\left|\sum_{i=1}^{n}w_{ij}\left\{y_{i} - \mu\left(\mathbf{w}_{i}^{T}\boldsymbol{\beta}\right) \right\} \right| \right. \\
\left. \leq \lambda + \sum_{r=1}^{\infty} \frac{\delta^{r} }{r!\sqrt{n} } \left\|\boldsymbol{\beta} \right\|_{1}^{r} \left\|\boldsymbol{\mu}^{\left(r\right)}\left(\mathbf{W}\boldsymbol{\beta} \right) \right\|_{2}\right],
\end{align*}
and
\begin{equation*}
\boldsymbol{\mu}^{\left(r\right)}\left(\mathbf{W}\boldsymbol{\beta} \right)  = \left(\mu^{\left(r\right)}\left(\mathbf{w}_{1}^{T}\boldsymbol{\beta} \right) , \dots, \mu^{\left(r\right)}\left(\mathbf{w}_{n}^{T}\boldsymbol{\beta} \right) \right)^{T}.
\end{equation*}
The proof is given in Appendix \ref{sec:ProofProp}.
\end{proposition}

The term 
\begin{align*}
\sum_{i=1}^{n}w_{ij}\left\{y_{i} - \mu\left(\mathbf{w}_{i}^{T}\boldsymbol{\beta}\right) \right\} 
\end{align*}
in the definition of $\Theta$ above, is the score function of the log-likelihood with respect to $\boldsymbol{\beta}$. In the classical ($p < n$) case, without measurement error, one would typically set $\lambda = \delta =0$, yielding the maximum likelihood estimate. Since $p>n$, and the maximum likelihood estimate is not well defined, we allow $\lambda>0$, to an extent which depends on the magnitude of the residuals $\mathbf{\epsilon}$. This means that the score differs from $0$ by at most $\lambda$. Furthermore, since we have covariate measurement error, we allow the score to be further away from zero, at a magnitude which depends on the complicated sum involving the regularization parameter $\delta$. Finally, we obtain a sparse estimate of the regression function by picking a sparse element of $\Theta$ according to some criterion. In analog with the GDS, we define the GMUS estimate as the element in $\Theta$ of minimum L1-norm. Using a finite number $R$ of terms in the Taylor expansion (\ref{eq:TaylorExpansion}), this yields the following definition of the GMUS:
\begin{align}\label{eq:GMUS}
&\hat{\boldsymbol{\beta}}_{MU}^{R} \in \text{arg}~\underset{\boldsymbol{\beta}}{\text{min}} \left\| \boldsymbol{\beta}\right\|_{1}, \text{ s.t. } \boldsymbol{\beta} \in \Theta^{R}, \text{ where} \\ \label{eq:ThetaGMUSR}
&\Theta^{R} =\left[ \boldsymbol{\beta} \in \mathbb{R}^{p}: \frac{1}{n}\underset{1\leq j \leq p}{\text{max}}\left|\sum_{i=1}^{n}w_{ij}\left\{y_{i} - \mu\left(\mathbf{w}_{i}^{T}\boldsymbol{\beta}\right) \right\} \right| \right.\\ \nonumber
 & \qquad \left.\leq \lambda + \sum_{r=1}^{R} \frac{\delta^{r} }{r! \sqrt{n}} \left\|\boldsymbol{\beta} \right\|_{1}^{r} \left\|\boldsymbol{\mu}^{\left(r\right)}\left(\mathbf{W}\boldsymbol{\beta} \right) \right\|_{2}\right].
\end{align}
In the rest of this paper, we use $R=1$, and we will show that this first order approximation yields good results in practice.

\subsection{Computation Using IRLS}
We suggest using a weighted least squares approach to compute $\hat{\boldsymbol{\beta}}_{MU}^{R}$. Extending the IRLS algorithm for the GDS \citep{James09}, we assume that $\mathbf{w}_{i}^{T}\boldsymbol{\beta}^{(k)}$ is the current estimate of the linear predictor for sample $i$ after completing the $k$th iteration. Now define the adjusted dependent covariate
\begin{equation}\label{eq:AdjustedOutcome}
z_{i} = \mathbf{w}_{i}^{T}\boldsymbol{\beta}^{(k)} + \frac{y_{i} - \mu\left( \mathbf{w}_{i}^{T}\boldsymbol{\beta}^{(k)}\right)}{\mu^{\prime}\left( \mathbf{w}_{i}^{T}\boldsymbol{\beta}^{(k)} \right)}, ~i=1,\dots,n.
\end{equation}
Next, we define a weight vector for each term in the Taylor expansion with $R$ terms,
\begin{equation}\label{eq:WeightVector}
\mathbf{V}^{(r)} = \left\{\mu^{(r)}\left( \mathbf{w}_{1}^{T}\boldsymbol{\beta}^{(k)} \right) , \dots, \mu^{(r)}\left(\mathbf{w}_{n}^{T}\boldsymbol{\beta}^{(k)} \right)\right\}^{T},
\end{equation}
for $r=1,\dots,R$, and introduce the matrix $\tilde{\mathbf{W}}\in \mathbb{R}^{n \times p}$ and the vector $\tilde{\mathbf{z}} \in \mathbb{R}^{n}$ with elements
\begin{equation}\label{eq:WeightedMatVec}
\tilde{w}_{ij} = \sqrt{V_{i}^{(1)}}w_{ij} \text{ and } \tilde{z}_{i} =\sqrt{V_{i}^{(1)}}z_{i},
\end{equation}
for $i=1,\dots,n$, $j=1,\dots,p$. The next iterate $\boldsymbol{\beta}^{(k+1)}$ is now given by
\begin{align}\label{eq:WeightedOptGMUS}
&\boldsymbol{\beta}^{(k+1)} \in \text{arg}~\underset{\boldsymbol{\beta}}{\text{min}} \left\| \boldsymbol{\beta}\right\|_{1}, \text{ s.t. }\\ \nonumber
& \frac{1}{n}\left\|\tilde{\mathbf{W}}^{T}\left(\tilde{\mathbf{z}} - \tilde{\mathbf{W}}\boldsymbol{\beta}\right) \right\|_{\infty} \leq \lambda + \sum_{r=1}^{R} \frac{\delta^{r} }{r! \sqrt{n}} \left\|\boldsymbol{\beta}\right\|_{1}^{r}\left\| \mathbf{V}^{(r)}\right\|_{2}
\end{align}
When we use a first order Taylor expansion ($R=1$), (\ref{eq:WeightedOptGMUS}) is equivalent to a linear program, as shown in Appendix \ref{sec:GMUSLinProg}.  The procedure is summarized in the following algorithm. As initial estimate $\boldsymbol{\beta}^{(0)}$, a convenient choice may be to take the GDS solution (\ref{eq:GDS}), corresponding to $\delta=0$. 

\begin{algorithmic}[1]
\REQUIRE Fix an inital estimate $\boldsymbol{\beta}^{(0)}$, $k=0$, $R \in \mathbb{N}$.
\REPEAT
\STATE Compute $\mathbf{V}^{(r)}$ according to (\ref{eq:WeightVector})
\STATE Compute $\mathbf{z}$ according to (\ref{eq:AdjustedOutcome})
\STATE Compute $\tilde{\mathbf{W}}$ and $\tilde{\mathbf{z}}$ according to (\ref{eq:WeightedMatVec})
\STATE Compute $\boldsymbol{\beta}^{(k+1)}$ by solving (\ref{eq:WeightedOptGMUS})
\STATE $k \leftarrow k+1$
\UNTIL{$\|\boldsymbol{\beta}^{(k)} - \boldsymbol{\beta}^{(k-1)}\| < \epsilon_{tol}$}
\RETURN $\hat{\boldsymbol{\beta}}_{MU}^{R} = \boldsymbol{\beta}^{(k)}$
\end{algorithmic}
On convergence, $\boldsymbol{\beta}^{(k+1)} = \boldsymbol{\beta}^{(k)}$ up to a small tolerance parameter $\epsilon_{tol}$, and $\boldsymbol{\beta}^{(k+1)}$ is solution of the GMUS (\ref{eq:GMUS}).

This algorithm is not guaranteed to converge, because the GMUS will generally have multiple solutions. In our simulation experiments with $n=200$ and $p=500$, we have experienced that it converges in a small number of iterations except when $\lambda$ and $\delta$ are close to zero. This is also demonstrated in Section \ref{sec:Convergence}. 

\section{A Lasso Analog of the GMUS}\label{sec:MUL}
\cite{Rosenbaum10} note that the convex optimization problem
\begin{align}\label{eq:MULasso}
\hat{\boldsymbol{\beta}}_{ML} \in \text{arg}~\underset{\boldsymbol{\beta}}{\text{min}}\frac{1}{2n} \left\| \mathbf{y} - \mathbf{W} \boldsymbol{\beta}\right\|_{2}^{2} 
+ \lambda \left\|\boldsymbol{\beta}\right\|_{1} + \frac{\delta}{2}\left\|\boldsymbol{\beta}\right\|_{1}^{2},
\end{align}
which we term the MU lasso (MUL), defines a lasso analog of the MUS. The solutions to (\ref{eq:MULasso}) are contained in the feasible set of the MUS, i.e.,
\begin{equation*}
\hat{\boldsymbol{\beta}}_{ML} \in \left\{ \boldsymbol{\beta}\in \mathbb{R}^{p} : \frac{1}{n} \left\| \mathbf{W}^{T}\left(\mathbf{y} - \mathbf{W}\boldsymbol{\beta} \right) \right\|_{\infty} \leq \lambda + \delta \left\|\boldsymbol{\beta}\right\|_{1}\right\},
\end{equation*}
which follows directly from the KKT conditions \citep{Rosenbaum10}. In a similar manner, we can define a generalized MUL (GMUL) by ensuring that its solution $\hat{\boldsymbol{\beta}}_{ML}^{R} \in \Theta^{R}$, where $\Theta^{R}$ is defined in (\ref{eq:ThetaGMUSR}). Define $\hat{\boldsymbol{\beta}}_{ML}^{R}$ as a vector which satisfies
\begin{align}\label{eq:GMULKKT}
&\frac{-1}{n}\sum_{i=1}^{n} w_{ij}\left\{ y_{i} - \mu\left( \mathbf{w}_{i}^{T}{\boldsymbol{\beta}}\right)\right\} = \\ \nonumber
&\qquad \tau_{j}\left\{\lambda + \sum_{r=1}^{R} \frac{\delta^{r} }{r!\sqrt{n}}  \left\| \boldsymbol{\beta}\right\|_{1}^{r} \left\|\boldsymbol{\mu}^{(r)}\left( \mathbf{W}\boldsymbol{\beta}\right) \right\|_{2} \right\},
\end{align}
where $|\tau_{j}| \leq 1$ and $\tau_{j} 1(\beta_{j}\neq 0) = \text{sign}(\beta_{j})$ for $j=1,\dots,p$ \citep{Buhlmann11}. 

We can formulate (\ref{eq:GMULKKT}) as a weighted least squares problem. Following (\ref{eq:AdjustedOutcome})-(\ref{eq:WeightedOptGMUS}), an iterate $\boldsymbol{\beta}^{(k+1)}$ given the present estimate $\boldsymbol{\beta}^{(k)}$ is a solution to
\begin{align}\label{eq:GMULIRLSKKT}
\frac{-1}{n}\tilde{\mathbf{w}}_{j}^{T} \left(\tilde{\mathbf{z}} - \tilde{\mathbf{W}} \boldsymbol{\beta} \right) = \tau_{j}\left\{ \lambda + \sum_{r=1}^{R} \gamma_{r}\left(r+1\right) \left\|\boldsymbol{\beta} \right\|_{1}^{r}\right\},
\end{align}
for $j=1,\dots,p$, where 
\begin{equation}
\gamma_{r} = \frac{\delta^{r} \left\|\mathbf{V}^{(r)} \right\|_{2}}{(r+1)! \sqrt{n}}, ~ r=1,\dots,R.
\label{eq:GMULGamma}
\end{equation}
However, (\ref{eq:GMULIRLSKKT}) are the KKT conditions corresponding to the convex optimization problem
\begin{align}\label{eq:GMULOptProb}
\text{min. } \frac{1}{2n} \left\| \tilde{\mathbf{z}} - \tilde{\mathbf{W}} \boldsymbol{\beta} \right\|_{2}^{2} + \lambda \left\| \boldsymbol{\beta}\right\|_{1} + \sum_{r=1}^{R} \gamma_{r}\left\|\boldsymbol{\beta} \right\|_{1}^{r+1}.
\end{align}
Hence, we can find the vector $\hat{\boldsymbol{\beta}}_{ML}^{R}$ satisfying (\ref{eq:GMULKKT}) by solving the optimization problem (\ref{eq:GMULOptProb}) in each step of an IRLS algorithm. The convexity of (\ref{eq:GMULOptProb}) follows from the fact that the composition of the convex L1 norm $\|\cdot\|_{1}$ and the convex and nondecreasing power function $(\cdot)^{r+1}$, $\|\cdot\|_{1}^{r+1}$, is itself a convex function \citep[p. 84]{Boyd04}. The solutions will satisfy $\hat{\boldsymbol{\beta}}_{ML}^{R} \in \Theta^{R}$, and the GMUL can thus be seen as a lasso analog of the GMUS. 

Unfortunately, due to the nonseparable penalty terms on the form\\ $\|\boldsymbol{\beta}\|_{1}^{r+1}$, even when $R=1$, (\ref{eq:GMULOptProb}) is not amenable to fast coordinatewise algorithms which have proven very useful for lasso-type problems \citep{Friedman2007, Friedman10}. Considering the $R=1$ case, it turns out that we can solve a lasso problem at each step of the IRLS algorithm, rather than the challenging problem (\ref{eq:GMULOptProb}). We start by noting that with $R=1$,  (\ref{eq:GMULOptProb}) is equivalent to
\begin{align*}
\text{min. } &\frac{-1}{n}\tilde{\mathbf{z}}^{T} \tilde{\mathbf{W}}\boldsymbol{\beta} + \boldsymbol{\beta}^{T}\left\{ \frac{1}{2n} \tilde{\mathbf{W}}^{T} \tilde{\mathbf{W}} + \gamma_{1} \mathbf{I}_{p} \right\}\boldsymbol{\beta}  \\
&\qquad +\lambda \left\| \boldsymbol{\beta}\right\|_{1} + \gamma_{1}\sum_{j=1}^{p} \sum_{l \neq j} \left| \beta_{j}\right| \left| \beta_{l}\right|.
\end{align*}
We can now replace the last penalty term in this expression with a weighted L1 penalty. The weight depends on the current estimate $\boldsymbol{\beta}^{(k)}$, and will hence be updated at each step of the IRLS algorithm. To be specific, given $\boldsymbol{\beta}^{(k)}$ we compute
\begin{align}\label{eq:GMULConverted}
 &\boldsymbol{\beta}^{(k+1)} \in \text{arg}~\underset{\boldsymbol{\beta}}{\text{min}} \bigg[\frac{-1}{n}\tilde{\mathbf{z}}^{T} \tilde{\mathbf{W}}\boldsymbol{\beta} +\\ \nonumber
& \boldsymbol{\beta}^{T}\left\{ \frac{1}{2n}  \tilde{\mathbf{W}}^{T} \tilde{\mathbf{W}} + \gamma_{1} \mathbf{I}_{p} \right\}\boldsymbol{\beta}   + \sum_{j=1}^{p} \omega_{j}^{(k)}  \left| \beta_{j}\right|\bigg] ,
\end{align}
where the weights are given by
\begin{align}
\omega_{j}^{(k)} = \lambda + \gamma_{1} \sum_{l \neq j} \left|\beta_{l}^{(k)}\right|, ~ j=1,\dots,p.
\label{eq:GMULOmega}
\end{align}
Since the Hessian $ (2n)^{-1}  \tilde{\mathbf{W}}^{T} \tilde{\mathbf{W}} + \gamma_{1} \mathbf{I}_{p}$ is always positive semidefinite for $\gamma_{1}\geq 0$, (\ref{eq:GMULConverted}) is an L1 constrained convex optimization problem, which can be efficiently solved with a coordinate descent algorithm. As we show in Appendix \ref{sec:CoorDescGMUL}, the coordinate-wise updates take the form
\begin{align}\label{eq:SoftThresholdLasso}
{\beta}_{j} \leftarrow \frac{S\left(\frac{1}{n} \sum_{i=1}^{n} \tilde{w}_{ij} \left( \tilde{z}_{i} - \sum_{l \neq j} \tilde{w}_{il} {\beta}_{l} \right), \omega_{j}^{(k)} \right)}{\frac{1}{n}\sum_{i=1}^{n} \tilde{w}_{ij}^{2} + 2 \gamma_{1}},
\end{align}
for $j=1,\dots,p,1,\dots$ until convergence, where $S(\cdot,\cdot)$ denotes the soft-thresholding operator
\begin{align}
S\left(a,b\right) = 
\begin{cases}
a-b, &\text{if } a > 0 \text{ and } b < \left|a\right| \\
a+b, &\text{if } a < 0 \text{ and } b < \left|a\right| \\
0, &\text{if } b \geq \left|a\right|.
\end{cases}
\label{eq:SoftThreshold}
\end{align}

We now have the following IRLS algorithm for computing  $\hat{\boldsymbol{\beta}}_{ML}^{R}$ with $R=1$:
\begin{algorithmic}[1]
\REQUIRE An inital estimate $\boldsymbol{\beta}^{(0)}$ exists, $k=1$.
\REPEAT
\STATE Compute $\mathbf{V}^{(1)}$ according to (\ref{eq:WeightVector})
\STATE Compute $\mathbf{z}$ according to (\ref{eq:AdjustedOutcome})
\STATE Compute $\tilde{\mathbf{W}}$ and $\tilde{\mathbf{z}}$ according to (\ref{eq:WeightedMatVec})
\STATE Compute $\gamma_{1}$ according to (\ref{eq:GMULGamma})
\STATE Compute $\omega_{j}^{(k)}$ according to (\ref{eq:GMULOmega}), $j=1,\dots,p$
\STATE Let $l=0$ and $\boldsymbol{\beta}^{(k+1,0)} \leftarrow \boldsymbol{\beta}^{(k)}$
\REPEAT
\FOR {$j=1,\dots,p$}
\STATE Update $\beta_{j}^{(k+1,l+1)}$ according to (\ref{eq:SoftThresholdLasso})
\ENDFOR
\STATE $l \leftarrow l+1$
\UNTIL{$\|\boldsymbol{\beta}^{(k+1,l+1)} - \boldsymbol{\beta}^{(k+1,l)}\| < \epsilon_{tol}$}
\RETURN $\boldsymbol{\beta}^{(k+1)} = \boldsymbol{\beta}^{(k+1,l+1)}$
\STATE $k \leftarrow k+1$
\UNTIL{$\|\boldsymbol{\beta}^{(k)} - \boldsymbol{\beta}^{(k-1)}\| < \epsilon_{tol}$}
\RETURN $\hat{\boldsymbol{\beta}}_{ML}^{R} = \boldsymbol{\beta}^{(k)}$
\end{algorithmic}

Again, the algorithm is not guaranteed to converge, because the solution is not in general unique, but we have experienced good convergence properties in practice as long as the constraint parameters $\lambda$ and $\delta$ are not too small. See also Section \ref{sec:Convergence}.

\section{Convergence of IRLS Algorithms}\label{sec:Convergence}
\begin{figure}%
\centering
\includegraphics[width=0.49\columnwidth]{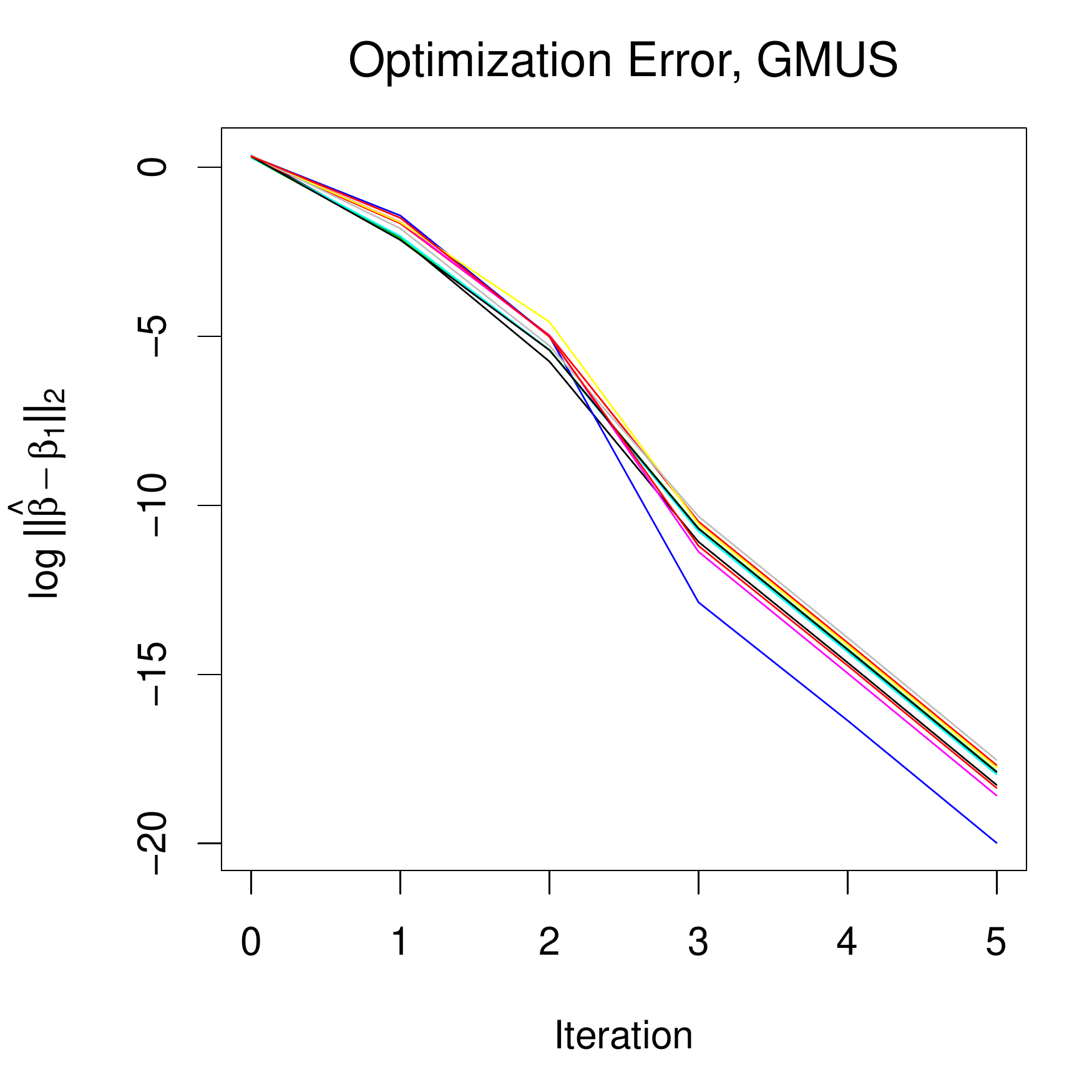}
\includegraphics[width=0.49\columnwidth]{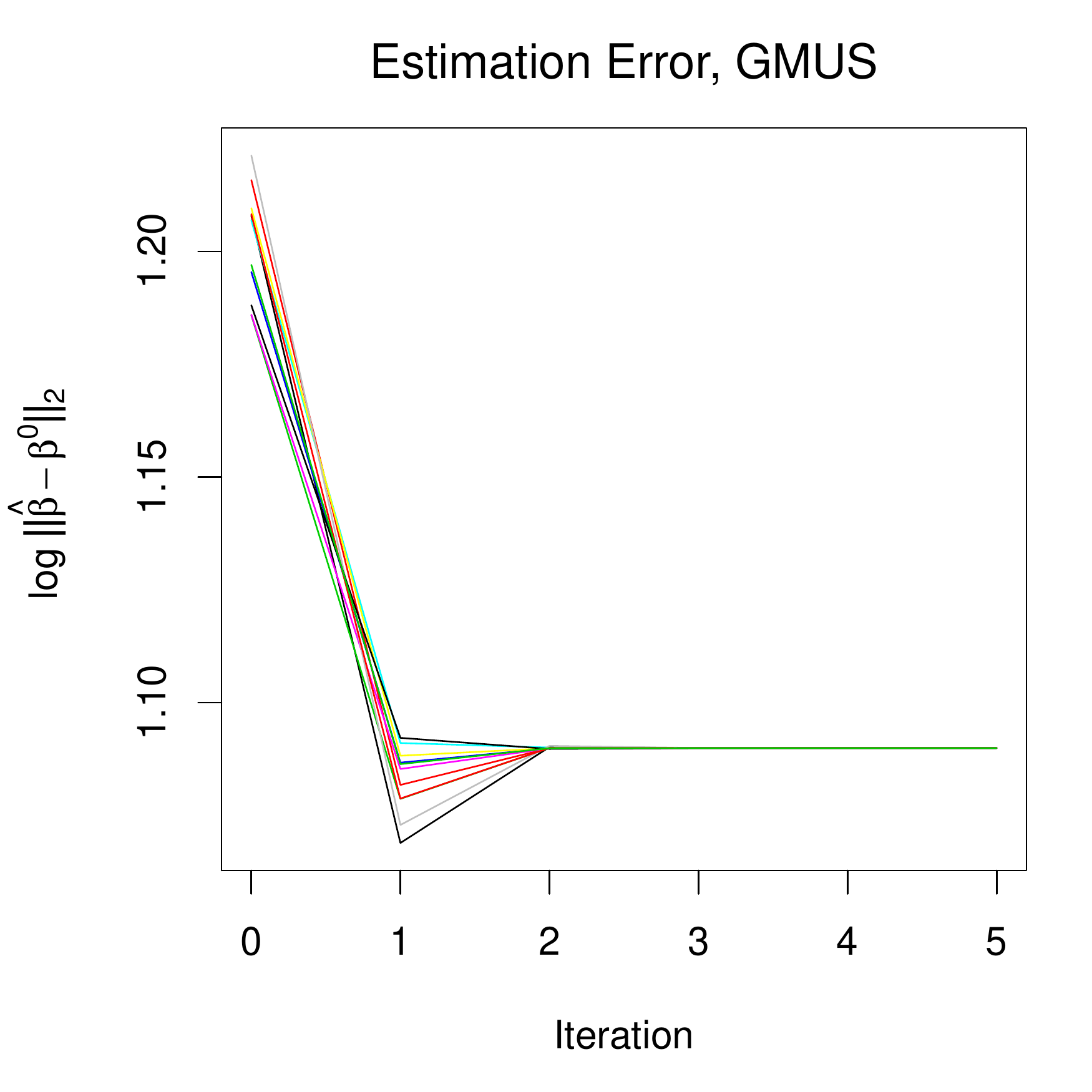}%
\caption{Convergence assessment of the IRLS algorithm for the GMUS.}%
\label{fig:GMUSdiagnostics}%
\end{figure}
We tested the convergence of the IRLS algorithm for the GMUS with $R=1$ using a technique similar to, e.g., \cite{Loh12}. We generated a problem instance with $n=200$, $p=500$, $s=10$, and $\boldsymbol{\beta}^{0}  = (1,\dots,1,0,\dots,0)^{T}$. The matrix of covariates had standard normally distributed entries $x_{ij}\sim N(0,1)$ and the measurement matrix had entries $w_{ij} = x_{ij} + u_{ij}$, with $u_{ij} \sim N(0,\sigma_{u})$, for $i=1,\dots,n$ and $j=1,\dots,p$, with $\sigma_{u}=0.2$. The response was binomially distributed with mean $(1+\exp\{-\mathbf{x}_{i}^{T}\boldsymbol{\beta}^{0}\})^{-1}$, and we set $\lambda=(1/3)\sqrt{{\log p}/n}$ and $\delta=0.1$. The value of $\lambda$ was chosen to be on the same order of magnitude as the theoretically optimal value, cf. \cite[p. 127]{Buhlmann11}. The GMUS with a logistic link function was computed $11$ times, each time with a random starting point for the IRLS algorithm. The left plot in Figure \ref{fig:GMUSdiagnostics} shows optimization errors of the last ten runs computed as the logarithm of the L2 distance between the current iterate and the corresponding estimate obtained in the first run. This gives a picture of the sensitivity to initial conditions, and we see that the difference between the estimates gets extremely small after very few iterations. The right plot in Figure \ref{fig:GMUSdiagnostics} shows the statistical error computed as the L2 distance between the current iterate and $\boldsymbol{\beta}^{0}$ for each of the 11 runs. Here we see that after two iterations of the IRLS algorithm, the statistical error is indistinguishable between the different runs. The algorithm converged very quickly in this case. We experienced the same good convergence properties in the simulation experiments of Section \ref{sec:Simulation} and on the microarray data of Section \ref{sec:Example}.

\begin{figure}%
\centering
\includegraphics[width=0.49\columnwidth]{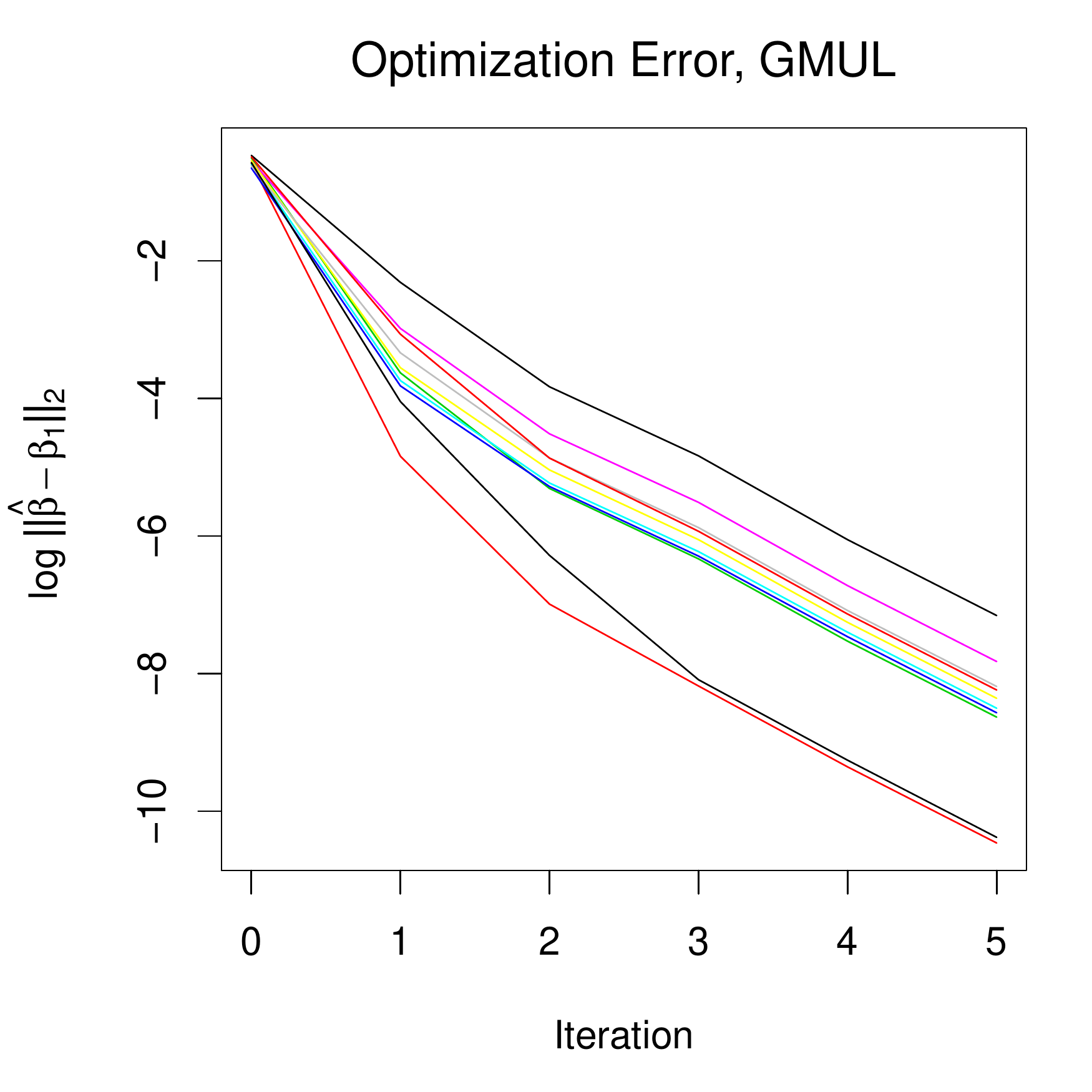}
\includegraphics[width=0.49\columnwidth]{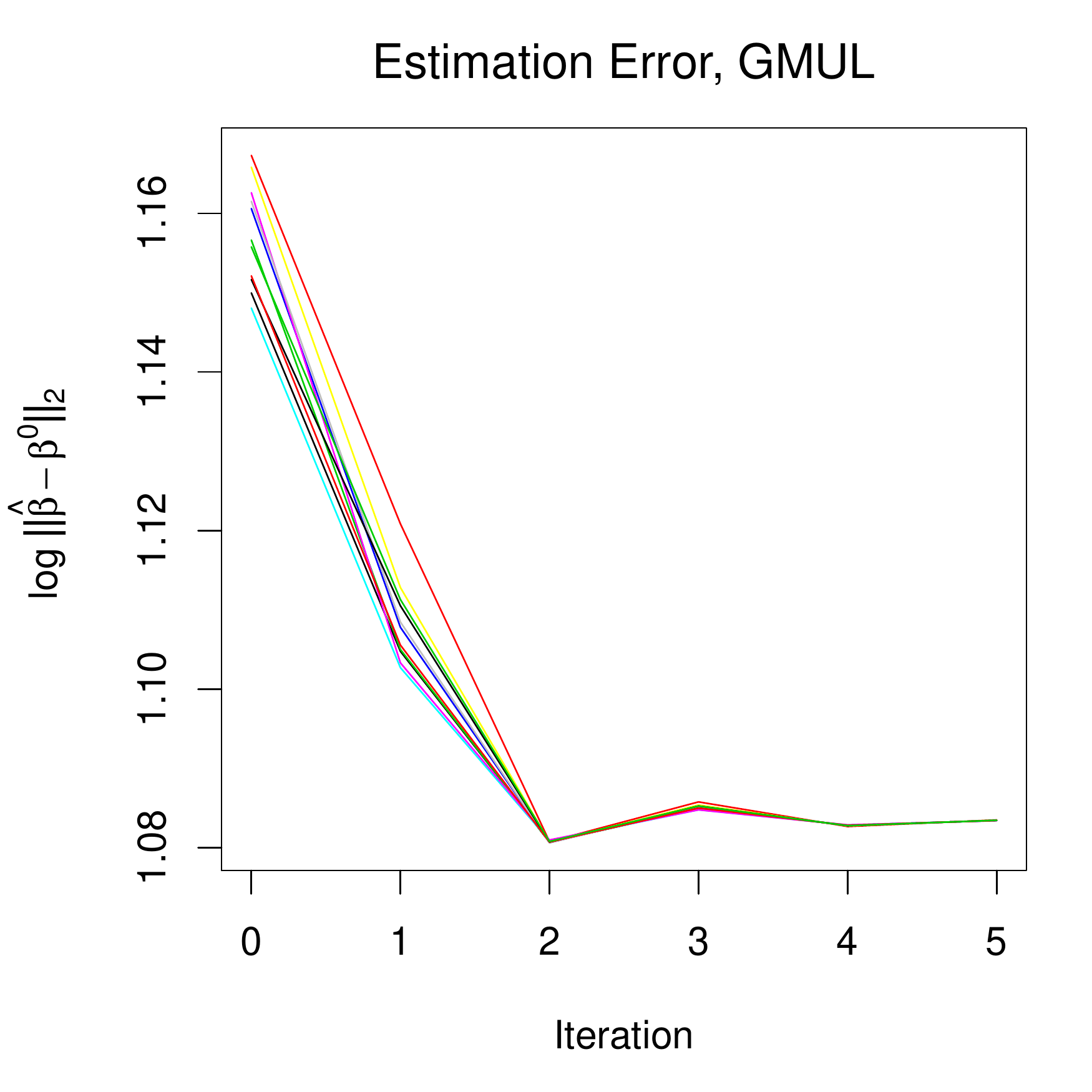}%
\caption{Convergence assessment of the IRLS algorithm for the GMUL.}%
\label{fig:GMULassodiagnostics}%
\end{figure}

The same experiment was done for the IRLS algorithm for the GMUL, and the results are shown in Figure \ref{fig:GMULassodiagnostics}. Also here we see that the algorithm converges with very few iterations. For the GMUL, the inner coordinate descent algorithm converged in less than ten steps in every iteration.

\section{Simulation Experiments}\label{sec:Simulation}
In this section we describe simulation experiments comparing the GMUL and the GMUS to the standard formulations of the GDS and the lasso. In all cases considered, we set $n=200$ and $s=10$. The matrix $\mathbf{X}$ had i.i.d. entries $x_{ij} \sim N(0,1)$, and conditional on $\mathbf{X}$, $\mathbf{W}$ had i.i.d. entries $w_{ij} = x_{ij} + u_{ij}$ with $u_{ij} \sim N(0,\sigma_{u})$, for $i=1,\dots,n$ and $j=1,\dots,p$. The response $y_{i}$ was either binomially distributed with mean $(1+\exp\{-\mathbf{x}_{i}^{T}\boldsymbol{\beta}^{0}\})^{-1}$ or Poisson distributed with mean $\exp(\mathbf{x}_{i}^{T}\boldsymbol{\beta}^{0})$. For each problem instance with data $(\mathbf{W}, \mathbf{y})$, two GDS and lasso fits were computed via ten-fold cross-validation: one fit corresponding to the minimum cross validated deviance, whose regularization parameter we denote $\hat{\lambda}_{min}$, and one fit corresponding to the largest regularization parameter within one standard error of the minimum, whose regularization parameter we denote $\hat{\lambda}_{se}$. The latter is suggested in \cite[p. 244]{Hastie09}. The solution to the GMUL was computed over a discrete grid of $\delta$ values, fixing $\lambda$ at the $\hat{\lambda}_{min}$ or $\hat{\lambda}_{se}$ obtained in the lasso fit. The same procedure was performed for the GMUS, using $\hat{\lambda}_{min}$ or $\hat{\lambda}_{se}$ from the GDS fit. For each setting, this procedure was repeated in $100$ independent Monte Carlo experiments.

Finally, the Monte Carlo average of the number of nonzero coefficients was plotted against the value of $\delta$, as shown in Figure \ref{fig:ElbowRuleLogistic}. According to the \emph{elbow rule} \citep{Rosenbaum10}, $\delta$ was chosen where the curve begins to flatten.  Hence, a common regularization parameter $\delta$ was chosen for all Monte Carlo simulations, while the $\lambda$ was chosen by cross-validation separately in each case. In Section \ref{sec:Example} we will show a real application, where $\delta$ is of course chosen specifically for the data set at hand.

The lasso solution was computed using the R package \verb!glmnet! \citep{Friedman10} and the GMUL was implemented in R and C++ using the \verb!Rcpp! package \citep{Eddelbuettel2011, Eddelbuettel2013}. The GDS and GMUS were implemented in R, utilizing the high-level interface to the GNU Linear Programming Kit provided by the package \verb!Rglpk! \citep{Theussl2013} to compute the solution to the linear program \eqref{eq:WeightedOptGMUS} in each step of the IRLS algorithm.

\subsection{Logistic Regression}
For the data with binomially distributed response, the logistic regression versions of the methods were used. The nonzero regression coefficients were set to $\beta_{j}^{0}=1$ for $j=1,\dots,s$, and the number of covariates was $p=500$. The measurement error standard deviation was either $\sigma_{u}=0.2$ or $\sigma_{u}=0.5$, respectively. 

\begin{figure}
\centering
\includegraphics[width=0.49\columnwidth]{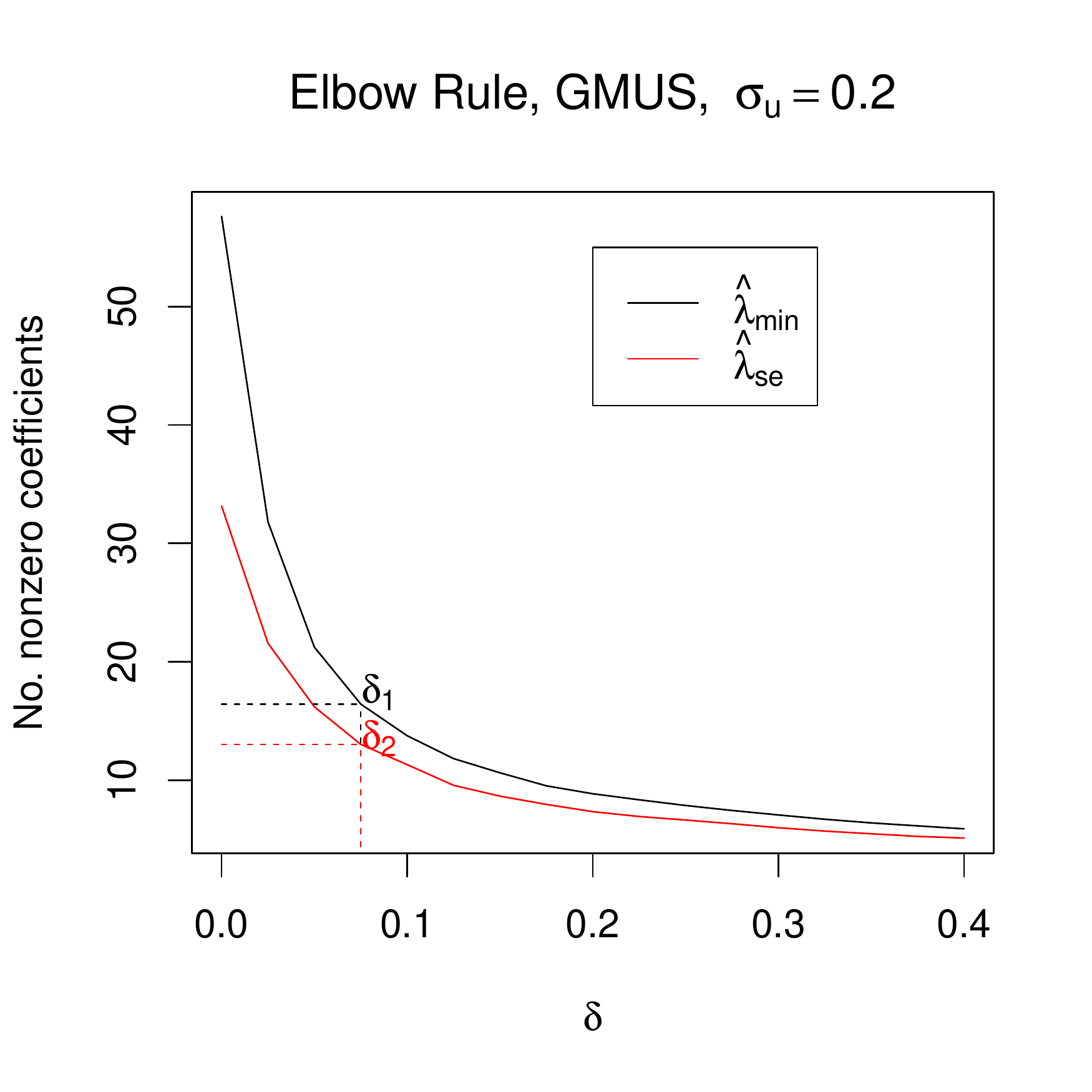}\\
\caption{Elbow rule with logistic regression. The average number of nonzero coefficients is plotted against $\delta$. $\hat{\lambda}_{min}$ denotes the value of $\lambda$ minimizing the cross-validation error for the GDS while $\hat{\lambda}_{se}$ denotes the largest $\lambda$ within one standard error of the minimum. $\delta=0$ corresponds to the GDS.}
\label{fig:ElbowRuleLogistic}
\end{figure}

Figure \ref{fig:ElbowRuleLogistic} shows the elbow plot for the GMUS and GDS in the $\sigma_{u}=0.2$ case. We see that the elbows are quite well defined. The elbow plots for all other settings are shown in the supplementary material.

\begin{table}
\renewcommand{\arraystretch}{1.3}
\caption{Results from the simulation experiment with logistic regression, with $p=500$ and $\sigma_{u}=0.2$. \#TP denotes the number of true positives and \#FP the number of false positives. Standard errors are shown in parentheses.}
\label{tab:LogisticSdu02}
\centering
\begin{tabular}{|l|lll|}
\hline
 & \#TP & \#FP  & \#TP/(\#TP + \#FP)\\
 \hline
 Lasso$(\hat{\lambda}_{min})$ &$9.64$ $(0.07)$ & $40.42$ $(1.59)$& $0.22$ $(0.01)$ \\
 GMUL$(\hat{\lambda}_{min}, \delta_{1})$ & $8.56$ $(0.17)$ & $10.14$ $(0.65)$& $0.50$ $(0.01)$ \\
  Lasso$(\hat{\lambda}_{se})$ &$8.93$ $(0.13)$ & $15.15$ $(0.91)$& $0.43$ $(0.02)$ \\
 GMUL$(\hat{\lambda}_{se}, \delta_{2})$ &$8.57$ $(0.15)$ & $8.24$ $(0.45)$& $0.55$ $(0.01)$  \\
 GDS$(\hat{\lambda}_{min})$ & $9.74$ $(0.05)$ & $47.87$ $(1.71)$& $0.19$ $(0.01)$\\
 GMUS$(\hat{\lambda}_{min}, \delta_{1})$ & $8.54$ $(0.11)$ & $7.89$ $(0.36)$& $0.54$ $(0.01)$\\
 GDS$(\hat{\lambda}_{se})$ &$9.30$ $(0.10)$ & $23.86$ $(1.30)$& $0.32$ $(0.01)$\\
GMUS$(\hat{\lambda}_{se}, \delta_{2})$ &$7.79$ $(0.13)$ & $5.24$ $(0.31)$& $0.63$ $(0.02)$\\
 \hline
\end{tabular}
\end{table}

Table \ref{tab:LogisticSdu02} shows the covariate selection properties of the different methods in the $\sigma_{u}=0.2$ case, with regularization parameters set as shown in Figure \ref{fig:ElbowRuleLogistic}. Overall, the GMUL and the GMUS display a drastic reduction in the number of false positives (FPs) compared to the GDS and the lasso, at the cost of a slight reduction in the number of true positives (TPs). This is also evident in the precision, the number of relevant covariates among those selected, shown in the rightmost column. While the lasso and GDS estimates computed have precision ranging from $0.19$ to $0.43$, the GMUL and the GMUS estimates have precision from $0.50$ up to $0.63$. Similar to the conclusion of \cite{Rosenbaum10} for linear models, Table \ref{tab:LogisticSdu02} shows that the standard formulations of the lasso and the GDS for logistic regression select too many covariates in the presence of measurement error. The GMUL and GMUS, which take the measurement error into account through the additional regularization parameter $\delta$, sharply improve the precision.

\begin{table}
\renewcommand{\arraystretch}{1.3}
\caption{Results from the simulation experiment with logistic regression, with $p=500$ and $\sigma_{u}=0.5$. \#TP denotes the number of true positives and \#FP the number of false positives. Standard errors are shown in parentheses.}
\label{tab:LogisticSdu05}
\centering
\begin{tabular}{|l|lll|}
\hline
& \#TP & \#FP & \#TP/(\#TP + \#FP) \\
 \hline
 Lasso$(\hat{\lambda}_{min})$ &$8.54$ $(0.16)$ & $31.68$ $(1.74)$& $0.26$ $(0.01)$  \\
 GMUL$(\hat{\lambda}_{min}, \delta_{1})$ & $7.42$ $(0.18)$ & $11.72$ $(0.68)$& $0.44$ $(0.02)$ \\
  Lasso$(\hat{\lambda}_{se})$ &$7.11$ $(0.22)$ & $11.63$ $(1.03)$& $0.49$ $(0.02)$ \\
 GMUL$(\hat{\lambda}_{se}, \delta_{2})$ & $6.47$ $(0.22)$ & $7.09$ $(0.61)$& $0.57$ $(0.02)$\\
 GDS$(\hat{\lambda}_{min})$ &$8.95$ $(0.14)$ & $43.45$ $(2.03)$& $0.20$ $(0.01)$\\
 GMUS$(\hat{\lambda}_{min}, \delta_{1})$ &$7.72$ $(0.17)$ & $9.73$ $(0.52)$& $0.48$ $(0.01)$\\
 GDS$(\hat{\lambda}_{se})$ &$8.04$ $(0.19)$ & $19.90$ $(1.50)$& $0.37$ $(0.02)$\\
GMUS$(\hat{\lambda}_{se}, \delta_{2})$ & $7.16$ $(0.18)$ & $7.76$ $(0.55)$& $0.54$ $(0.02)$\\
 \hline
\end{tabular}
\end{table}

Table \ref{tab:LogisticSdu05} shows the similar results with $\sigma_{u}=0.5$. This is a harder problem, as the measurement error is rather large, but also in this case the GMUL and the GMUS reduce the number of FPs considerably. 

\subsection{Poisson Regression}
For the data with Poisson distributed response, the Poisson regression versions of the methods were used. The nonzero regression coefficients were set to $\beta_{j}^{0}=0.2$ for $j=1,\dots,s$, and the measurement error standard deviation was $\sigma_{u}=0.2$.

\begin{table}
\renewcommand{\arraystretch}{1.3}
\caption{Results from the simulation experiment with Poisson regression, with $p=150$ and $\sigma_{u}=0.2$. \#TP denotes the number of true positives and \#FP the number of false positives. Standard errors are shown in parentheses.}
\label{tab:PoissonP150}
\centering
\begin{tabular}{|l|lll|}
\hline
 & \#TP & \#FP & \#TP/(\#TP + \#FP) \\
 \hline
 Lasso$(\hat{\lambda}_{min})$ &  $9.97$ $(0.02)$ & $28.09$ $(1.07)$& $0.28$ $(0.01)$\\
 GMUL$(\hat{\lambda}_{min}, \delta_{1})$ & $9.51$ $(0.06)$ & $6.97$ $(0.30)$& $0.59$ $(0.01)$\\
  Lasso$(\hat{\lambda}_{se})$ & $9.65$ $(0.07)$ & $10.64$ $(0.65)$& $0.52$ $(0.02)$\\
 GMUL$(\hat{\lambda}_{se}, \delta_{2})$ & $9.37$ $(0.09)$ & $6.48$ $(0.41)$& $0.63$ $(0.01)$\\
 GDS$(\hat{\lambda}_{min})$ &$9.95$ $(0.02)$ & $43.44$ $(1.54)$& $0.20$ $(0.01)$\\
 GMUS$(\hat{\lambda}_{min}, \delta_{1})$ &$9.34$ $(0.09)$ & $12.12$ $(0.53)$& $0.46$ $(0.01)$\\
 GDS$(\hat{\lambda}_{se})$ &$9.44$ $(0.09)$ & $16.87$ $(1.06)$& $0.41$ $(0.01)$\\
GMUS$(\hat{\lambda}_{se}, \delta_{2})$ &$7.71$ $(0.16)$ & $3.51$ $(0.29)$& $0.73$ $(0.02)$ \\
 \hline
\end{tabular}
\end{table}

Since the exponential mean function for Poisson regression is more highly nonlinear than the mean function for logistic regression, one might suspect that the first order Taylor approximation considered in this paper yields a poorer fit for Poisson regression. We therefore started with $p=150$ covariates, creating a somewhat easier problem. Table \ref{tab:PoissonP150} shows the results, from which it is clear that the standard lasso and GDS select too many covariates. First of all, the GDS and lasso with $\lambda = \hat{\lambda}_{min}$ select a large number of false positives. Secondly, the GMUL and the GMUS with $\lambda = \hat{\lambda}_{min}$ perform better than the GDS and lasso with $\lambda = \hat{\lambda}_{se}$. For example, the GMUL with $\lambda=\hat{\lambda}_{min}$ selects on average $9.51$ out of the $10$ TPs and $6.97$ FPs. The lasso with $\lambda = \hat{\lambda}_{se}$, on the other hand, selects $9.65$ TPs and $10.64$ FPs. Hence, while the number of TPs selected is almost identical, the lasso selects on average $3.67$ more FPs than the GMUS in this case. The corresponding numbers for the GMUS are $9.34$ TPs and $12.12$ FPs, versus $9.44$ TPs and $16.87$ FPs of the GDS.

\begin{table}
\renewcommand{\arraystretch}{1.3}
\caption{Results from the simulation experiment with Poisson regression, with $p=500$ and $\sigma_{u}=0.2$. \#TP denotes the number of true positives and \#FP the number of false positives. Standard errors are shown in parentheses.}
\label{tab:PoissonP500}
\centering
\begin{tabular}{|l|lll|}
\hline
 & \#TP & \#FP & \#TP/(\#TP + \#FP) \\
 \hline
 Lasso$(\hat{\lambda}_{min})$ &  $9.80$ $(0.05)$ & $42.54$ $(1.94)$& $0.21$ $(0.01)$\\
 GMUL$(\hat{\lambda}_{min}, \delta_{1})$ & $9.14$ $(0.10)$ & $13.94$ $(0.56)$& $0.42$ $(0.01)$\\
  Lasso$(\hat{\lambda}_{se})$ & $8.23$ $(0.23)$ & $12.00$ $(0.81)$& $0.41^*$\\
 GMUL$(\hat{\lambda}_{se}, \delta_{2})$ & $7.97$ $(0.23)$ & $8.27$ $(0.54)$& $0.49^*$\\
 GDS$(\hat{\lambda}_{min})$ &$9.58$ $(0.12)$ & $57.96$ $(2.15)$& $0.14^*$\\
 GMUS$(\hat{\lambda}_{min}, \delta_{1})$ &$8.01$ $(0.15)$ & $12.60$ $(0.54)$& $0.39^*$\\
 GDS$(\hat{\lambda}_{se})$ &$8.24$ $(0.19)$ & $22.88$ $(1.64)$& $0.26^*$\\
GMUS$(\hat{\lambda}_{se}, \delta_{2})$ &$7.03$ $(0.20)$ & $8.41$ $(0.67)$& $0.46^*$\\
 \hline
\end{tabular}
\end{table}

Table \ref{tab:PoissonP500} shows the corresponding results with $p=500$. For some of the methods, there were cases in which no covariate was selected. In these cases, the average precision and its standard error are not well defined, and we report instead the average number of TPs divided by the average number of selected covariates, and mark the corresponding cells with an asterisk. That is, the numbers marked with an asterisk in the fourth column were computed by dividing the number in the second column in the same row by the sum of the numbers in the second and third columns in the same row. The GMUS and GMUL have consistently better covariate selection performance than the lasso and the GDS. For example, comparing Lasso$(\hat{\lambda}_{min})$ and  GMUL$(\hat{\lambda}_{min}, \delta_{1})$, we see that adding the additional regularization $\delta_{1}$ after cross-validation doubles the precision, by reducing the average number of irrelevant covariates selected by $28.6$. 

\section{Analysis of a Microarray Data Set}\label{sec:Example}
We now show an application to a data set containing $22,815$ normalized microarray gene expression measurements as well as measurements of bone mineral density (BMD), for $84$ Norwegian women \citep{Reppe10}. As microarray measurements are known to be noisy and subject to various sources of bias \citep{Boulesteix08,Rocke01,Tadesse05}, this context is very appropriate for the GMUL and the GMUS. The total hip T-score BMD was used as outcome variable, and the subjects were classified as having BMD below median ($y=1$) or above median ($y=0$).

\begin{figure}
\centering
\includegraphics[width=0.49\columnwidth]{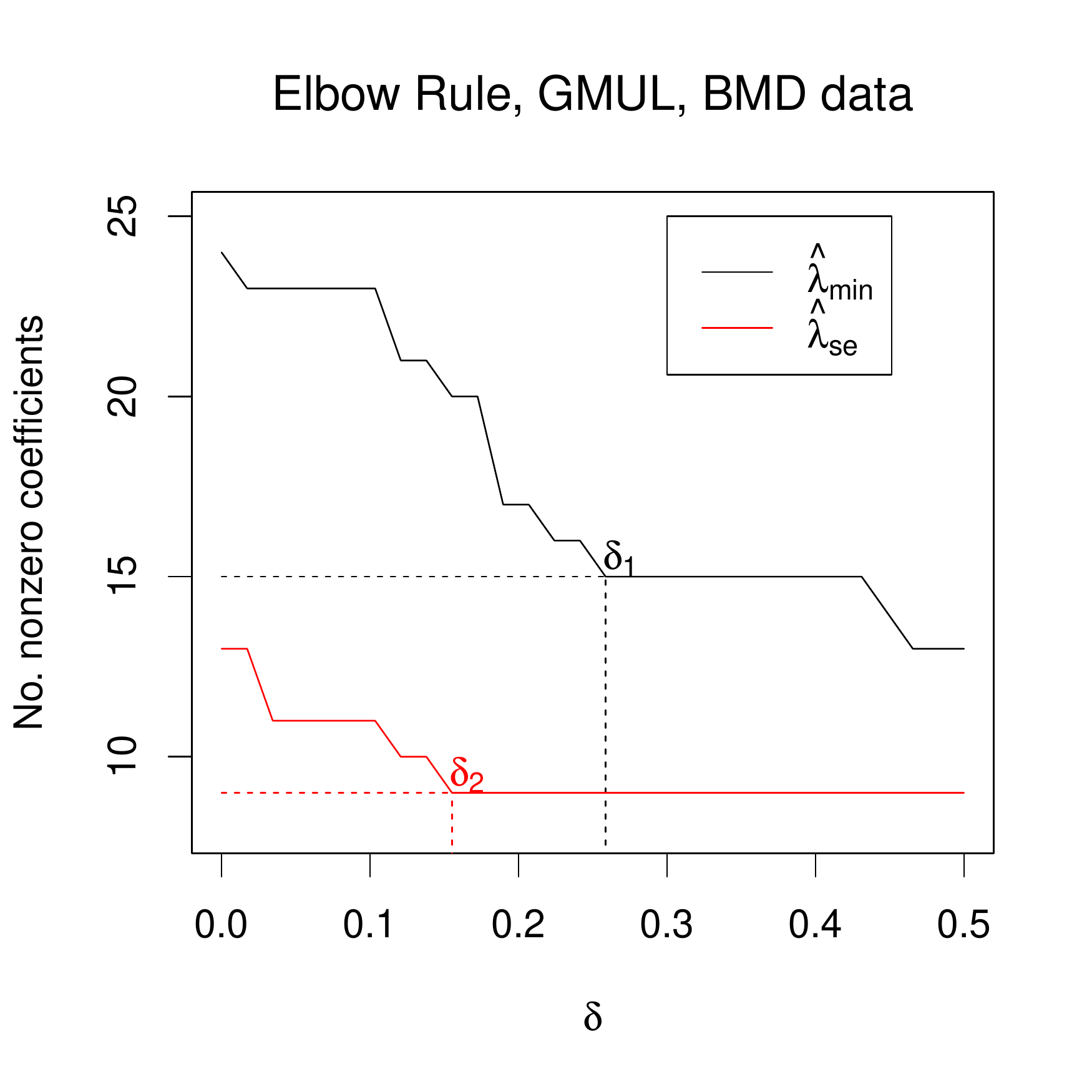}
\caption{Elbow rule for the BMD example.}
\label{fig:ElbowBMDGMUL}
\end{figure}

Our IRLS algorithm, with an inner coordinate descent loop for computing the GMUL estimate, scales very well with the number of covariates $p$, and we were therefore able to compute GMUL estimates for the full dataset with $p=22,815$. The current implementation of the GMUS, on the other hand, is based on using the simplex method to solve the linear program (\ref{eq:WeightedOptGMUS}) in each step of the IRLS algorithm. Doing this repeatedly over a grid of $\lambda$ and $\delta$ values, turned out to be computationally infeasible when $p$ is this large. We thus only report results obtained using the lasso and GMUL. Developing a tailor-made solver for (\ref{eq:WeightedOptGMUS}) using the interior-point method \citep{Boyd04} could overcome this problem, but is beyond the scope of this paper.

As in the simulation experiments, we computed two standard lasso fits of the BMD data, using \verb!glmnet! \citep{Friedman10}: One corresponding to the minimum cross-validated deviance, with regularization parameter $\hat{\lambda}_{min}$, and one corresponding to $\hat{\lambda}_{se}$, the largest regularization parameter within one standard error of the minimum cross-validated deviance. Starting at these two $\lambda$ values, we gradually increased $\delta$ from $0$ on a discrete grid with spacing $0.025$. Figure \ref{fig:ElbowBMDGMUL} shows the elbow plots for these two GMUL estimators. Both curves have a long range of $\delta$ values for which the number of nonzero coefficients is constant, and we choose the corresponding regularization parameters $\delta_{1}$ and $\delta_{2}$ at the minimum values over these regions.

\begin{table}
\renewcommand{\arraystretch}{1.3}
\caption{Gene symbols and standardized regression coefficients of genes selected by lasso and GMUL with $\lambda = \hat{\lambda}_{min}$.}
\label{tab:LassoMinGenes}
\centering
\begin{tabular}{lrlr}
\multicolumn{2}{c}{Lasso$(\hat{\lambda}_{min})$} & \multicolumn{2}{c}{GMUL$(\hat{\lambda}_{min}, \delta_{1})$} \\
  \hline
 Gene symbol & Coef. & Probe ID & Coef. \\ 
  \hline
MIR22HG & 0.36 & MIR22HG & 0.23 \\ 
 FAM118B & 0.32 & FAM118B & 0.20 \\ 
 ITGBL1 & -0.22 & BE551142 & -0.18 \\ 
 (control seq.) & -0.21 & (control seq.) & -0.15 \\ 
 BE551142 & -0.20 & ADI1 & 0.12 \\ 
 FKBP5 & 0.16 & POLR2D & -0.09 \\ 
 ADI1 & 0.15 & ITGBL1 & -0.09 \\ 
Hs.380027 & -0.14 & PCGF1 & -0.09 \\ 
 POLR2D & -0.12 & SOST & -0.09 \\ 
 PCGF1 & -0.10 & COPS4 & 0.06 \\ 
RPH3A & -0.10 & FKBP5 & 0.04 \\ 
COPS4 & 0.10 & ELTD1 & 0.04 \\ 
Hs.633128 & -0.09 & CARD8 & -0.02 \\ 
TSC22D1-AS1 & -0.08 & Hs.633128 & -0.02 \\ 
CARD8 & -0.08 &  &  \\ 
COL11A1 & -0.08 &  &  \\ 
AA463449 & -0.08 &  &  \\ 
CCDC169 & -0.07 &  &  \\ 
SOST & -0.06 &  &  \\ 
AF086063 & -0.06 &  &  \\ 
MON1B & -0.06 &  &  \\ 
DYNLRB1 & -0.05 &  &  \\ 
ELTD1 & 0.04 &  & \\ 
SEMA3F & 0.00 &  & \\ 
   \hline
\end{tabular}
\end{table}

Table \ref{tab:LassoMinGenes} shows the gene symbols and standardized regression coefficients for the lasso estimate with $\lambda = \hat{\lambda}_{min}$ and the GMUL with $\lambda = \hat{\lambda}_{min}$ and $\delta = \delta_{1}$. The lasso selected $23$ genes and one control sequence, and the GMUL reduced this number to $13+1$, but all genes selected by the GMUL were also selected by the lasso. The GMUL hence worked by removing genes from the selected set, without adding any new ones. Based on our simulation experiments, it seems plausible that the genes selected by the lasso but not by the GMUL were indeed false positives.

\begin{table}
\renewcommand{\arraystretch}{1.3}
\caption{Gene symbols and standardized regression coefficients of genes selected by lasso and GMUL with $\lambda = \hat{\lambda}_{se}$.}
\label{tab:Lasso1seGenes}
\centering
\begin{tabular}{lrlr}
\multicolumn{2}{c}{Lasso$(\hat{\lambda}_{min})$} & \multicolumn{2}{c}{GMUL$(\hat{\lambda}_{min}, \delta_{1})$} \\
  \hline
Gene symbol & Coef. & Gene symbol & Coef. \\ 
  \hline
MIR22HG & 0.26 & MIR22HG & 0.21 \\ 
  FAM118B & 0.23 & FAM118B & 0.19 \\ 
  BE551142 & -0.18 & BE551142 & -0.14 \\ 
  (control seq.) & -0.14 & (control seq.) & -0.12 \\ 
  POLR2D & -0.10 & POLR2D & -0.08 \\ 
  ADI1 & 0.09 & ADI1 & 0.05 \\ 
  SOST & -0.07 & SOST & -0.05 \\ 
  PCGF1 & -0.07 & PCGF1 & -0.04 \\ 
  ITGBL1 & -0.04 &  &  \\ 
  COPS4 & 0.02 &  &  \\ 
  ELTD1 & 0.01 &  &  \\ 
  CARD8 & -0.01 &  & \\ 
  Hs.633128 & -0.01 &  &  \\ 
   \hline
\end{tabular}
\end{table}

Table \ref{tab:Lasso1seGenes} shows the corresponding table when using $\lambda = \hat{\lambda}_{se}$. Here, the lasso selected $12$ genes and one control sequence, all of which were also chosen by the lasso with $\lambda = \hat{\lambda}_{min}$. The GMUL removed $5$ genes, ending up with $7+1$ selected covariates. Again, the genes selected by the GMUL were a subset of the genes selected by the lasso. The genes selected by the GMUL with $\lambda = \hat{\lambda}_{se}$ were also a subset of the genes selected by the GMUL with $\lambda = \hat{\lambda}_{min}$.

\section{Conclusion}\label{sec:Conclusion}
This paper focuses on covariate selection in high-dimensional GLMs when the covariates are subject to measurement error. We generalize the MUS \citep{Rosenbaum10}, which is limited to linear models, by considering an $R$th order Taylor approximation of the GLM mean function. Furthermore, we develop the GMUL, a lasso analog of the GMUS. By not requiring an estimate of the measurement error covariance matrix, the proposed methods give an important practical advantage. 

Both the GMUS and the GMUL can be computed using IRLS. For computational reasons, we only consider Taylor approximations of order $R=1$. In this case, the GMUS requires solving a linear program at each step of the IRLS algorithm, while the GMUL can be computed with an inner coordinate descent loop at each step of the IRLS algorithm. We demonstrate in simulation experiments with logistic and Poisson regression, that the standard formulations of the lasso and GDS select a large number of false positives when the covariates are subject to measurement error, and that the GMUS and the GMUL can reduce this number while mainly keeping the true positives. 

As the main focus of this paper is covariate selection in the presence of measurement error, the simulation results in Section \ref{sec:Simulation} focused on the number of TPs, FPs, and the precision. Another important aspect is the estimation error. We have observed in practice that the GMUL and the GMUS tend to have a smaller $\ell_{1}$ estimation error than the lasso and GDS, and a larger $\ell_{2}$ estimation error. This can be explained by the fact that the GMUL and GMUS have better covariate selection properties (lower $\ell_{1}$ error), and penalize more (higher $\ell_{2}$ error). If a low $\ell_{2}$ estimation error is of main interest, it may therefore be wiser to use the bias correction proposed by \cite{Loh12} or \cite{Rosenbaum13}.

In our current implementations, the GMUL can efficiently handle $p$ on the order of tens of thousands, while the linear program (\ref{eq:WeightedOptGMUS}) solved in each step of the IRLS algorithm for the GMUS becomes increasingly slow when $p$ is larger than about a thousand. An interesting problem for further study is therefore to develop an algorithm for (\ref{eq:WeightedOptGMUS}) which scales better with $p$. In addition, our current implementations are restricted to first order Taylor approximations. Although this has been shown to work well, it would be interesting to develop algorithms for computing the GMUL and the GMUS using higher order approximations.

Although not considered in this paper, the GMUL and the GMUS can also be used in problems with missing data. This is particularly relevant when imputation methods are computationally infeasible due to the large number of covariates.


%

\appendix
\section{Proof of Proposition \ref{prop:FeasibleSet}}\label{sec:ProofProp}
It follows from the Taylor series expansion \eqref{eq:TaylorExpansion} that
\begin{equation*}
\mu\left(\mathbf{w}_{i}^{T}\boldsymbol{\beta}^{0}\right) = \mu\left(\mathbf{x}_{i}^{T}\boldsymbol{\beta}^{0}\right) - \sum_{r=1}^{\infty} \frac{\mu^{(r)}\left( \mathbf{w}_{i}^{T}\boldsymbol{\beta}^{0}\right)}{r!} \left(-\mathbf{u}_{i}^{T}\boldsymbol{\beta}^{0}\right)^{r}.
\end{equation*}
This gives, for $j=1,\dots,p$,
\begin{align*}
&\frac{1}{n}\left|\sum_{i=1}^{n} w_{ij}\left(y_{i} - \mu\left(\mathbf{w}_{i}^{T}\boldsymbol{\beta}^{0} \right) \right) \right| = \\
&\frac{1}{n}\left|\sum_{i=1}^{n} w_{ij}\left(\epsilon_{i}+ \sum_{r=1}^{\infty} \frac{\mu^{(r)}\left( \mathbf{w}_{i}^{T}\boldsymbol{\beta}^{0}\right)}{r!} \left(-\mathbf{u}_{i}^{T}\boldsymbol{\beta}^{0}\right)^{r}\right) \right|\leq\\
&\frac{1}{n}\left|\sum_{i=1}^{n} w_{ij}\epsilon_{i} \right| + \frac{1}{n}\left|\sum_{i=1}^{n} w_{ij} \sum_{r=1}^{\infty} \frac{\mu^{(r)}\left( \mathbf{w}_{i}^{T}\boldsymbol{\beta}^{0}\right)}{r!} \left(-\mathbf{u}_{i}^{T}\boldsymbol{\beta}^{0}\right)^{r} \right| \leq \\
&\lambda + \frac{1}{n}\left|\sum_{i=1}^{n} w_{ij} \sum_{r=1}^{\infty} \frac{\mu^{(r)}\left( \mathbf{w}_{i}^{T}\boldsymbol{\beta}^{0}\right)}{r!} \left(-\mathbf{u}_{i}^{T}\boldsymbol{\beta}^{0}\right)^{r} \right| \leq \\
&\lambda + \frac{1}{n}\sum_{i=1}^{n}\left| w_{ij} \sum_{r=1}^{\infty} \frac{\mu^{(r)}\left( \mathbf{w}_{i}^{T}\boldsymbol{\beta}^{0}\right)}{r!} \left(-\mathbf{u}_{i}^{T}\boldsymbol{\beta}^{0}\right)^{r} \right|,
\end{align*}
where we inserted $\epsilon_{i} = y_{i} - \mu\left(\mathbf{w}_{i}^{T}\boldsymbol{\beta}^{0}\right)$ in the first step, we used the triangle inequality in the second step, we inserted the left bound in \eqref{eq:MUBounds} in the third step, and finally used the generalized triangle inequality. Next, we have
\begin{align*}
&\frac{1}{n}\sum_{i=1}^{n}\left| w_{ij} \sum_{r=1}^{\infty} \frac{\mu^{(r)}\left( \mathbf{w}_{i}^{T}\boldsymbol{\beta}^{0}\right)}{r!} \left(-\mathbf{u}_{i}^{T}\boldsymbol{\beta}^{0}\right)^{r} \right| \leq \\
& \frac{1}{n}\left(\sum_{i=1}^{n} w_{ij}^{2} \right)^{\frac{1}{2}} \left(\sum_{i=1}^{n} \left\{ \sum_{r=1}^{\infty} \frac{\mu^{(r)}\left( \mathbf{w}_{i}^{T}\boldsymbol{\beta}^{0}\right)}{r!} \left(-\mathbf{u}_{i}^{T}\boldsymbol{\beta}^{0}\right)^{r}\right\}^{2} \right)^{\frac{1}{2}}  \\
&= \frac{1}{\sqrt{n}} \left(\sum_{i=1}^{n} \left\{ \sum_{r=1}^{\infty} \frac{\mu^{(r)}\left( \mathbf{w}_{i}^{T}\boldsymbol{\beta}^{0}\right)}{r!} \left(-\mathbf{u}_{i}^{T}\boldsymbol{\beta}^{0}\right)^{r}\right\}^{2} \right)^{\frac{1}{2}},
\end{align*}
where we used Hölder's inequality in the first step and the assumption \eqref{eq:StandardizedCovs} that the covariates are standardized to have mean zero and unit variance in the second step. We now note that the last term above is the L2 norm $\|\mathbf{v}\|_{2}$ of a vector $\mathbf{v} \in \mathbb{R}^{n}$ with elements
\begin{equation*}
v_{i} = \sum_{r=1}^{\infty} \frac{\mu^{(r)}\left( \mathbf{w}_{i}^{T}\boldsymbol{\beta}^{0}\right)}{r!} \left(-\mathbf{u}_{i}^{T}\boldsymbol{\beta}^{0}\right)^{r},\quad i=1,\dots,n.
\end{equation*}
We thus have
\begin{align*}
&\frac{1}{\sqrt{n}} \left(\sum_{i=1}^{n} \left\{ \sum_{r=1}^{\infty} \frac{\mu^{(r)}\left( \mathbf{w}_{i}^{T}\boldsymbol{\beta}^{0}\right)}{r!} \left(-\mathbf{u}_{i}^{T}\boldsymbol{\beta}^{0}\right)^{r}\right\}^{2} \right)^{\frac{1}{2}} \leq \\
&\frac{1}{\sqrt{n}} \sum_{r=1}^{\infty} \left( \sum_{i=1}^{n}\left\{\frac{\mu^{(r)}\left( \mathbf{w}_{i}^{T}\boldsymbol{\beta}^{0}\right)}{r!} \right\}^{2}\left(-\mathbf{u}_{i}^{T}\boldsymbol{\beta}^{0}\right)^{2r} \right)^{\frac{1}{2}} \\
&\frac{1}{\sqrt{n}} \sum_{r=1}^{\infty} \left( \sum_{i=1}^{n}\left\{\frac{\mu^{(r)}\left( \mathbf{w}_{i}^{T}\boldsymbol{\beta}^{0}\right)}{r!} \right\}^{2} \left\|\mathbf{u}_{i}\right\|_{\infty}^{2r} \left\|\boldsymbol{\beta}\right\|_{1}^{2r}\right)^{\frac{1}{2}} \leq \\
&\sum_{r=1}^{\infty}\frac{\delta^{r}\left\|\boldsymbol{\beta}\right\|_{1}^{r}}{r!\sqrt{n}}  \left( \sum_{i=1}^{n}\left\{\mu^{(r)}\left( \mathbf{w}_{i}^{T}\boldsymbol{\beta}^{0}\right) \right\}^{2}  \right)^{\frac{1}{2}}= \\
&  \sum_{r=1}^{\infty}\frac{\delta^{r}\left\|\boldsymbol{\beta}\right\|_{1}^{r}}{r!\sqrt{n}} \left\|\boldsymbol{\mu}^{(r)}\left(\mathbf{W}\boldsymbol{\beta}^{0} \right) \right\|_{2} ,
\end{align*}
where we used the triangle inequality in the first step, Höffding's inequality in the second step, and finally used the right bound in \eqref{eq:MUBounds} in the second last step.

Putting the pieces together, it follows that
\begin{align*}
\frac{1}{n}\left|\sum_{i=1}^{n} w_{ij}\left(y_{i} - \mu\left(\mathbf{w}_{i}^{T}\boldsymbol{\beta}^{0} \right) \right) \right| \leq 
 \lambda +\sum_{r=1}^{\infty}\frac{\delta^{r}}{r!\sqrt{n}} \left\|\boldsymbol{\beta}^{0}\right\|_{1}^{r} \left\|\boldsymbol{\mu}^{(r)}\left(\mathbf{W}\boldsymbol{\beta}^{0} \right) \right\|_{2} 
\end{align*}
for $j=1,\dots,p$, which proves that $\boldsymbol{\beta}^{0} \in \Theta$.

\section{Computing the Solution to \eqref{eq:WeightedOptGMUS}}\label{sec:GMUSLinProg}
We can simplify the computation of \eqref{eq:WeightedOptGMUS} by introducing the auxiliary variable $\mathbf{u}\in \mathbb{R}^{p}$. We then get the equivalent problem
\begin{align*}
&\text{mininimize } \mathbf{1}_{p}^{T} \mathbf{u} \text{ (with respect to } \mathbf{u}, \boldsymbol{\beta} \text{)}\\
&\text{subject to } -\mathbf{u} \leq \boldsymbol{\beta} \leq \mathbf{u}, \\
&-\sum_{r=1}^{R} \frac{\delta^{r}}{r! \sqrt{n}} \left(\mathbf{1}_{p}^{T} \mathbf{u}\right)^{r}\left\|\mathbf{V}^{(r)}\right\|_{2} \mathbf{1}_{p}+ \frac{1}{n}\tilde{\mathbf{W}}^{T}\tilde{\mathbf{W}} \boldsymbol{\beta} \leq \lambda\mathbf{1}_{p} + \frac{1}{n}\tilde{\mathbf{W}}^{T} \tilde{\mathbf{z}},\\
&\text{and}\\
&-\sum_{r=1}^{R} \frac{\delta^{r}}{r! \sqrt{n}} \left(\mathbf{1}_{p}^{T} \mathbf{u}\right)^{r}\left\|\mathbf{V}^{(r)}\right\|_{2}\mathbf{1}_{p}- \frac{1}{n}\tilde{\mathbf{W}}^{T}\tilde{\mathbf{W}} \boldsymbol{\beta} \leq \lambda \mathbf{1}_{p}- \frac{1}{n}\tilde{\mathbf{W}}^{T} \tilde{\mathbf{z}}.
\end{align*}
When $R=1$, the problem \eqref{eq:WeightedOptGMUS} is thus equivalent to the linear program
\begin{align*}
&\text{mininimize } \mathbf{1}_{p}^{T} \mathbf{u}\text{ (with respect to } \mathbf{u}, \boldsymbol{\beta} \text{)} \\
&\text{subject to } -\mathbf{u} \leq \boldsymbol{\beta} \leq \mathbf{u}, \\
&- \frac{\delta}{\sqrt{n}} \mathbf{1}_{p}^{T} \mathbf{u}\left\|\mathbf{V}^{(1)}\right\|_{2}\mathbf{1}_{p}+ \frac{1}{n}\tilde{\mathbf{W}}^{T}\tilde{\mathbf{W}} \boldsymbol{\beta} \leq  \lambda \mathbf{1}_{p}+ \frac{1}{n}\tilde{\mathbf{W}}^{T} \tilde{\mathbf{z}},\\
&\text{and}\\
&-\frac{\delta}{\sqrt{n}} \mathbf{1}_{p}^{T} \mathbf{u}\left\|\mathbf{V}^{(1)}\right\|_{2}\mathbf{1}_{p}- \frac{1}{n}\tilde{\mathbf{W}}^{T}\tilde{\mathbf{W}} \boldsymbol{\beta} \leq  \lambda \mathbf{1}_{p}- \frac{1}{n}\tilde{\mathbf{W}}^{T} \tilde{\mathbf{z}},
\end{align*}
which can be solved by standard software.

\section{Coordinate Descent Algorithm for GMUL}\label{sec:CoorDescGMUL}
We describe here the coordinate descent algorithm used to solve (\ref{eq:GMULConverted}). Our goal is to find a $\boldsymbol{\beta}$ minimizing the function
\begin{align*}
f\left(\boldsymbol{\beta} \right) = -n^{-1}\tilde{\mathbf{z}}^{T} \tilde{\mathbf{W}}\boldsymbol{\beta} + \boldsymbol{\beta}^{T}\left\{ \left(2n\right)^{-1}  \tilde{\mathbf{W}}^{T} \tilde{\mathbf{W}} + \gamma_{1} \mathbf{I}_{p} \right\}\boldsymbol{\beta}   + \sum_{j=1}^{p} \omega_{j}^{(k)} \left|\beta_{j} \right|,
\end{align*}
which can be written equivalently as
\begin{align*}
f\left(\boldsymbol{\beta} \right) = &-n^{-1} \sum_{i=1}^{n} \tilde{z}_{i} \sum_{j=1}^{p} \tilde{w}_{ij} \beta_{j} + \left(2n\right)^{-1} \sum_{i=1}^{n} \left( \sum_{j=1}^{p} \tilde{w}_{ij} \beta_{j}\right)^{2} \\
& + \gamma_{1} \sum_{j=1}^{p} \beta_{j}^{2} +  \sum_{j=1}^{p} \omega_{j}^{(k)} \left|\beta_{j} \right|.
\end{align*}
The partial derivatives of $f\left(\boldsymbol{\beta}\right)$ with respect to  $\beta_{j}$, $j=1\dots,p$, can be written as
\begin{align*}
\frac{\partial f}{\partial \beta_{j}} = &-n^{-1} \sum_{i=1}^{n} \tilde{z}_{i} \tilde{w}_{ij} + n^{-1} \sum_{i=1}^{n} \tilde{w}_{ij} \sum_{l\neq j} \tilde{w}_{il}\beta_{l} \\
& +n^{-1} \beta_{j} \sum_{i=1}^{n} \tilde{w}_{ij}^{2} +2\gamma_{1} \beta_{j} + \omega_{j}^{(k)} \tau_{j},
\end{align*}
where $\tau_{j} = 1$ if $\beta_{j} > 0$, $\tau_{j} = -1$ if $\beta_{j} < 0$, and $\tau_{j} \in [-1,1]$ if $\beta_{j} = 0$. Setting $\partial f / \partial \beta_{j} = 0$, we find the analytical solution to (\ref{eq:GMULConverted}),
\begin{align*}
\beta_{j} = \frac{n^{-1} \sum_{i=1}^{n}\tilde{w}_{ij} \left(\tilde{z}_{i}   - \sum_{l \neq j} \tilde{w}_{ij} \beta_{l} \right) - \omega_{j}^{\left(k\right)} \tau_{j}}{ n^{-1} \sum_{i=1}^{n} \tilde{w}_{ij}^{2} + 2\gamma_{1}}
\end{align*}
for $j=1,\dots,p$. Since $\boldsymbol{\tau}$ is implicitly defined, we compute $\boldsymbol{\beta}$ iteratively using the coordinate descent updates
\begin{align*}
\hat{\beta}_{j} \leftarrow \frac{S\left(n^{-1}\sum_{i=1}^{n} \tilde{w}_{ij} \left(\tilde{z}_{i} - \sum_{l \neq j} \tilde{w}_{il} \hat{\beta}_{l} \right), \omega_{j}^{(k)}\right)}{\frac{1}{n} \sum_{i=1}^{n} \tilde{w}_{ij}^{2} + 2 \gamma_{1}}, 
\end{align*}
for $j=1,\dots,p,1,\dots$ until convergence \citep{Friedman2007}. $S(a,b)$ is the soft-thresholding operator (\ref{eq:SoftThreshold}). On convergence, we set $\boldsymbol{\beta}^{(k+1)} = \hat{\boldsymbol{\beta}}$. 
\section*{Acknowledgment}

The authors would like to thank Bin Yu for discussions and Sjur Reppe for providing the bone density data.



\bibliographystyle{chicago}
\bibliography{references}

\end{document}